\documentclass[prl,aps,twocolumn,superscriptaddress,showpacs,showkeys,amsmath,amssymb,floatfix,noshowpacs,noshowkeys]{revtex4}
\usepackage{booktabs,graphicx,verbatim}
\usepackage[colorlinks=true,citecolor=blue,filecolor=blue,linkcolor=blue,urlcolor=blue,pdftex]{hyperref}
\newcommand{\1}[1]{\, \mathrm{#1}} 

\newcommand{\arxiv}[1]{\href{http://arxiv.org/abs/#1}{\texttt{arXiv:#1}}}

\begin{document}

\title{Exclusion of Leptophilic Dark Matter Models \\ using XENON100 Electronic Recoil Data}

\newcommand{\bern}{\affiliation{Albert Einstein Center for Fundamental Physics, University of Bern, Bern, Switzerland}}
\newcommand{\bologna}{\affiliation{Department of Physics and Astrophysics, University of Bologna and INFN-Bologna, Bologna, Italy}}
\newcommand{\coimbra}{\affiliation{Department of Physics, University of Coimbra, Coimbra, Portugal}}
\newcommand{\columbia}{\affiliation{Physics Department, Columbia University, New York, NY, USA}}
\newcommand{\lngs}{\affiliation{INFN-Laboratori Nazionali del Gran Sasso and Gran Sasso Science Institute, L'Aquila, Italy}}
\newcommand{\mainz}{\affiliation{Institut f\"ur Physik \& Exzellenzcluster PRISMA, Johannes Gutenberg-Universit\"at Mainz, Mainz, Germany}}
\newcommand{\heidelberg}{\affiliation{Max-Planck-Institut f\"ur Kernphysik, Heidelberg, Germany}}
\newcommand{\munster}{\affiliation{Institut f\"ur Kernphysik, Wilhelms-Universit\"at M\"unster, M\"unster, Germany}}
\newcommand{\nikhef}{\affiliation{Nikhef and the University of Amsterdam, Science Park, Amsterdam, Netherlands}}
\newcommand{\nyuad}{\affiliation{New York University Abu Dhabi, Abu Dhabi, United Arab Emirates}}
\newcommand{\purdue}{\affiliation{Department of Physics and Astronomy, Purdue University, West Lafayette, IN, USA}}
\newcommand{\rensselaer}{\affiliation{Department of Physics, Applied Physics and Astronomy, Rensselaer Polytechnic Institute, Troy, NY, USA}}
\newcommand{\rice}{\affiliation{Department of Physics and Astronomy, Rice University, Houston, TX, USA}}
\newcommand{\shanghai}{\affiliation{Department of Physics \& Astronomy, Shanghai Jiao Tong University, Shanghai, China}}
\newcommand{\subatech}{\affiliation{SUBATECH, Ecole des Mines de Nantes, CNRS/In2p3, Universit\'e de Nantes, Nantes, France}}
\newcommand{\torino}{\affiliation{INFN-Torino and Osservatorio Astrofisico di Torino, Torino, Italy}}
\newcommand{\ucla}{\affiliation{Physics \& Astronomy Department, University of California, Los Angeles, CA, USA}}
\newcommand{\weizmann}{\affiliation{Department of Particle Physics and Astrophysics, Weizmann Institute of Science, Rehovot, Israel}}
\newcommand{\zurich}{\affiliation{Physik-Institut, University of Zurich, Zurich, Switzerland}}

\author{E.~Aprile}\columbia
\author{F.~Agostini}\lngs\bologna
\author{M.~Alfonsi}\nikhef
\author{L.~Arazi}\weizmann
\author{K.~Arisaka}\ucla
\author{F.~Arneodo}\nyuad
\author{M.~Auger}\zurich
\author{C.~Balan}\coimbra
\author{P.~Barrow}\zurich
\author{L.~Baudis}\zurich
% ...

\collaboration{The XENON Collaboration}\noaffiliation

\begin{abstract}
Laboratory experiments searching for galactic dark matter particles scattering off nuclei have so far not been able to establish a discovery. We use data from the XENON100 experiment to search for dark matter interacting with electrons. With no evidence for a signal above the low background of our experiment, we exclude a variety of representative dark matter models that would induce electronic recoils. For axial-vector couplings to electrons, we exclude cross-sections above $6\times10^{-35}\1{cm^2}$ for particle masses of $m_\chi = 2\1{GeV/c^2}$. Independent of the dark matter halo, we exclude leptophilic models as explanation for the long-standing DAMA/LIBRA signal, such as couplings to electrons through axial-vector interactions at a~$4.4\sigma$ confidence level, mirror dark matter at~$3.6\sigma$, and luminous dark matter at~$4.6\sigma$.
\end{abstract}

\maketitle

Dark Matter in the form of Weakly Interacting Massive Particles (WIMPs) is typically expected to induce nuclear recoils in a terrestrial detector target~\cite{Bergstrom:2012fi} with an annually modulated rate due to the motion of the Earth around the Sun~\cite{Drukier:1986tm,Freese:2012xd}. Although such a modulation has been observed by the DAMA/LIBRA collaboration using sodium iodine~\cite{Bernabei:2013xsa}, it is difficult to interpret it as a dark matter signal, given the null results from other experiments~\cite{Savage:2008er,Akerib:2013tjd,Agnese:2014aze}. In fact, dark matter-induced nuclear recoils are excluded by these results unless one invokes models that are fine-tuned to create a signal only in DAMA/LIBRA but not in other experiments~\cite{Bai:2009cd,Chang:2010en,Chang:2010pr}. In contrast, dark matter-induced electronic recoils appear as a viable explanation for the observed modulation because exclusions of other experiments do not apply directly in this case~\cite{Kopp:2009et,Chang:2014tea,Bell:2014tta}. We use data from the XENON100 detector to rule out this possibility for three different, representative dark matter models.

\begin{figure}[htbp]
\centering\includegraphics[width=1\columnwidth,clip=true,trim=0pt 0pt 0pt 0pt]{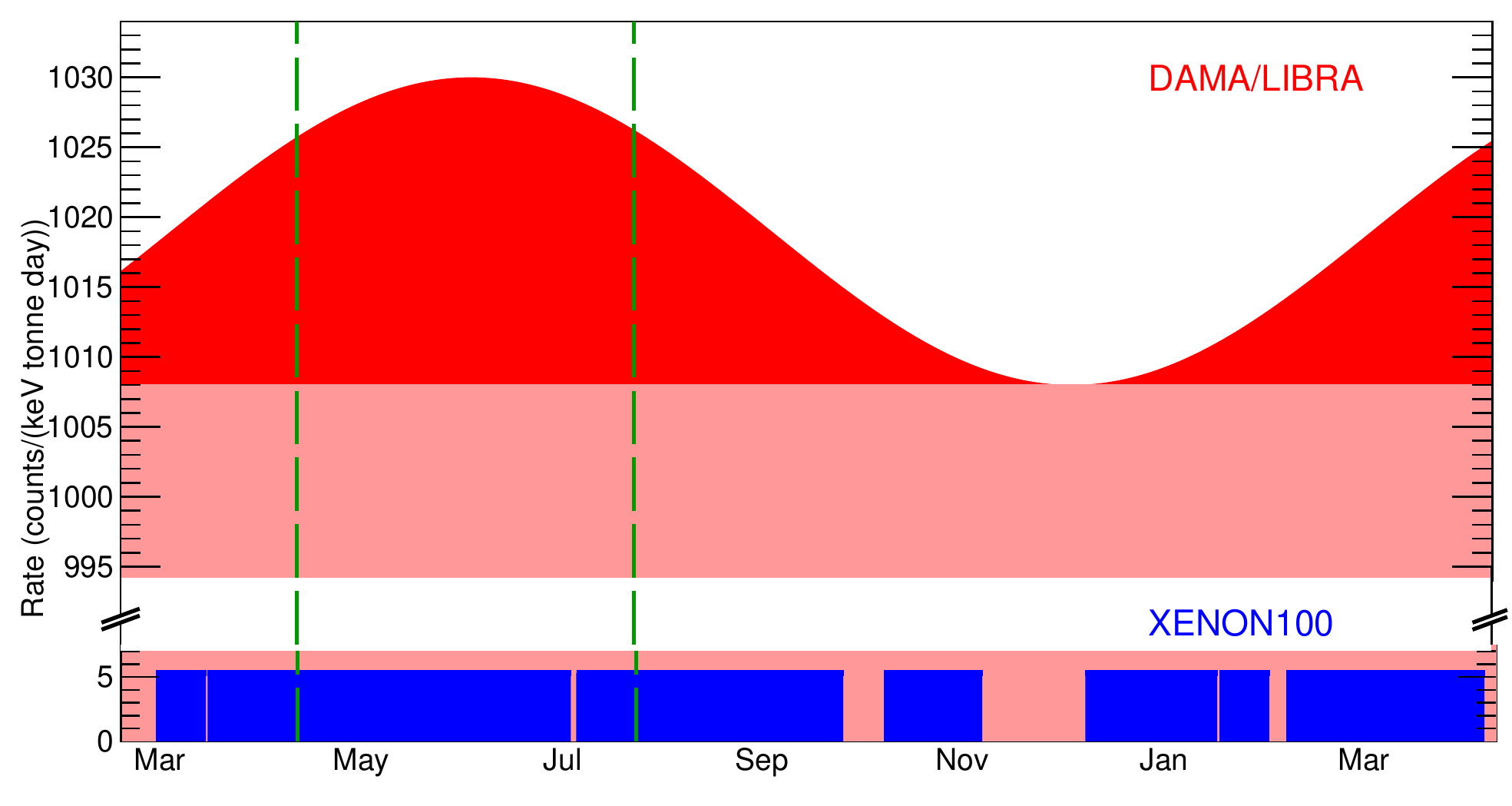}
\caption{Fig.~1. \textbf{Conceptual illustration of the analysis.} Shown is the DAMA/LIBRA rate (red)~\cite{Bernabei:2008yi} with the modulated rate in $(2-6)\1{keV}$ from the fit parameters in~\cite{Bernabei:2013xsa} (dark red). The distribution of the XENON100 live time (blue) is indicated with its average background rate of $5.3\1{events/(keV\cdot tonne\cdot day)}$, which shows dents due to maintenance or calibration campaigns. The region between the dashed lines (green) indicates the 70~summer live days where the modulated signal is expected to be largest.}
\end{figure}

We interpret data from the XENON100 detector that were acquired between February 28, 2011 and March 31, 2012 for a total exposure of 224.6 live days and $34\1{kg}$ fiducial mass. We have previously searched this data set for spin-independent~\cite{Aprile:2012nq} and spin-dependent~\cite{Aprile:2013doa} WIMP-induced nuclear recoils as well as for axion-induced electronic recoils~\cite{Aprile:2014eoa}. XENON100 is located in the Gran Sasso underground laboratory. It consists of a liquid xenon target that is operated as a low-background time projection chamber~\cite{Aprile:2011dd}. Each particle interaction results in two signals: The prompt scintillation signal (S1) is used here for energy estimation, and the delayed ionization signal (S2) allows for 3D~vertex reconstruction. Data reduction is performed in order to select single-scatter low-energy ($<10\1{keV}$) recoils in the fiducial volume, while retaining maximal detector efficiency~\cite{Aprile:2012vw,Aprile:2014eoa}. At low energies, the remaining background of XENON100 is dominated by forward-scattered Compton events, resulting in a flat spectrum with a rate of $5.3\1{events/(keV\cdot tonne\cdot day)}$ in the fiducial volume~\cite{Aprile:2011vb} (File~A1). This rate is more than two orders of magnitude lower than the average background rate of about $1019\1{events/(keV\cdot tonne\cdot day)}$ reported by DAMA/LIBRA in the same energy interval~\cite{Bernabei:2008yi,Kudryavtsev:2010a}, and even smaller than their reported annual modulation amplitude of $(11.2\pm1.2)\1{events/(keV\cdot tonne\cdot day)}$~\cite{Bernabei:2013xsa}. Because the DAMA/LIBRA collaboration has not published the composition of their background at low energies, we test the minimum dark matter signal that would be required to cause the observed modulation. In this scenario, the constant spectrum is fully attributed to background, and only the modulated part itself is attributed to a 100\% modulated dark matter signal as illustrated in (Fig.~1). We ignore the practical difficulties of realizing such a highly modulated signal~\cite{HerreroGarcia:2011aa,Freese:2012xd} but conservatively consider it as the case that is most challenging to exclude. The dark matter-induced rate would then be zero on December~2nd, and twice the measured modulation amplitude on June~2nd. It follows that there is an optimized time interval to consider for best sensitivity. To find this interval, the signal expected in XENON100 was simulated for different time intervals centered around June~2nd. We take into account uncertainties from counting statistics in XENON100 and DAMA/LIBRA, as well as the systematic uncertainty from the conversion of keV energy into S1~\cite{Aprile:2014eoa}. The optimum time interval is found to be 70~live days around June~2nd, roughly corresponding to April 2011--August 2011 (Fig.~1) as indicated. Our expected sensitivity varies by less than $0.1\sigma$ with changes of this interval of $\pm 40$~live days. A dedicated analysis of the time stability of XENON100 electron recoil data will be presented elsewhere~\cite{Aprile:2015ac}.

\textbf{WIMP axial-vector coupling to electrons:}
A relativistic treatment of dark matter-electron scattering shows that keV-scale electronic recoils can only be induced by dark matter particles with masses $m_{\chi}\gtrsim1\1{GeV/c^2}$ scattering inelastically off electrons with momenta on the order of MeV/c~\cite{Kopp:2009et,Bernabei:2007gr}.A qualitatively similar result is obtained by a simple non-relativistic treatment of elastic two-body scattering. As shown in~\cite{Kopp:2009et}, even if the dark matter has tree-level (first-order) interactions only with leptons, loop-induced dark matter-hadron interactions dominate the experimental signatures and make the usual exclusions based on nuclear recoil analyses applicable. Thus, we consider here axial-vector $\vec{A}\otimes\vec{A}$ couplings between dark matter and leptons, since in this case, loop contributions vanish, while the WIMP-electron coupling is not suppressed by additional small factors of velocity $v$ or mass ratio $m_e/m_{\chi}$.

\begin{figure}[htbp]
\centering\includegraphics[width=1\columnwidth,clip=true,trim=0pt 0pt 0pt 120pt]{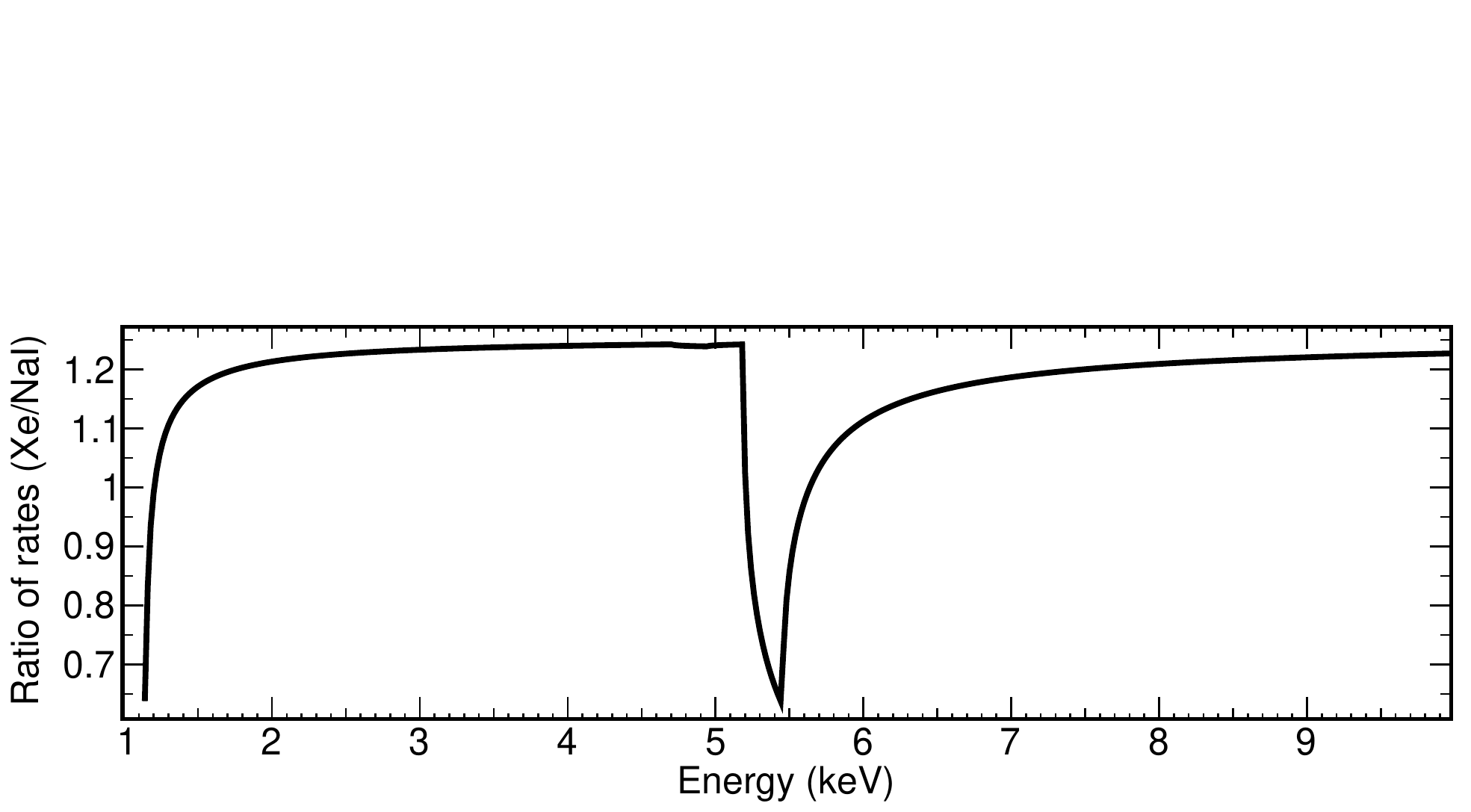}
\caption{Fig.~2. \textbf{Calculated ratio of the differential rates} in xenon and sodium iodide for inelastic WIMP-electron scattering through axial-vector coupling. The structures around 1 and 5~keV are owing to the small difference in the binding energies of the $3s$ and $2s$ shells in xenon and iodine.}
\end{figure}

We use equation~(30) in~\cite{Kopp:2009et}, with an additional factor of~2 to account for electron occupancy from spin, to calculate the differential rate for WIMP-electron scattering (File~A2). The expected rate includes a sum over the atomic shells of the target, and for each shell, integrates the momentum wave function of the electrons to get the contribution at a given recoil energy. Given the requirement that the energy deposited in the detector must be more than the binding energy of the electron, the largest contribution to the rate in a sodium iodide target comes from the $3s$ shell of iodine. The contributions from sodium are two orders of magnitude smaller. The momentum-space wave functions for xenon atoms and iodine anions are nearly identical as a result of their similar electron structure. This has the important consequence that a comparison between sodium iodide and xenon is independent of the dark matter halo. The ratio of the calculated differential rates in xenon and sodium iodide are shown in Fig.~2 as a function of deposited energy, considering the full shell structure. This ratio has negligible dependence on the WIMP mass.

We contrast the DAMA/LIBRA signal, interpreted as WIMPs coupling to electrons through axial-vector interactions, with XENON100 data. The energy spectrum of the modulation amplitude~\cite{Bernabei:2013xsa} is multiplied by the energy-dependent ratio from Fig.~2 and by a constant factor of 1.88, which accounts for the time integral of the modulated signal that is expected in our 70~summer live days (Fig.~1). The deposited electronic recoil energy in XENON100 is estimated from the S1 signal, measured in photoelectrons (PE), using the NESTv0.98 model~\cite{Szydagis:2011tk} which consistently fits the available data~\cite{Manalaysay:2009yq,Aprile:2012an,Baudis:2013cca,Szydagis:2013sih}. The energy scale, shown in~\cite{Aprile:2014eoa}, includes a systematic uncertainty that decreases from 20\% to 7\% from 1~keV to 10~keV, reflecting the spread and uncertainties in the measurements. The S1 generation is modelled as a Poisson process and the PMT resolution is taken into account in order to obtain the predicted XENON100 S1 spectrum from the scaled energy spectrum~\cite{Aprile:2012vw}. Our resolution is a factor 2 worse than that of DAMA/LIBRA; the feature at 5.2~keV in Fig.~2 is lost in this process.

\begin{figure}[htbp]
\centering\includegraphics[width=1\columnwidth,clip=true,trim=0pt 0pt 0pt 0pt]{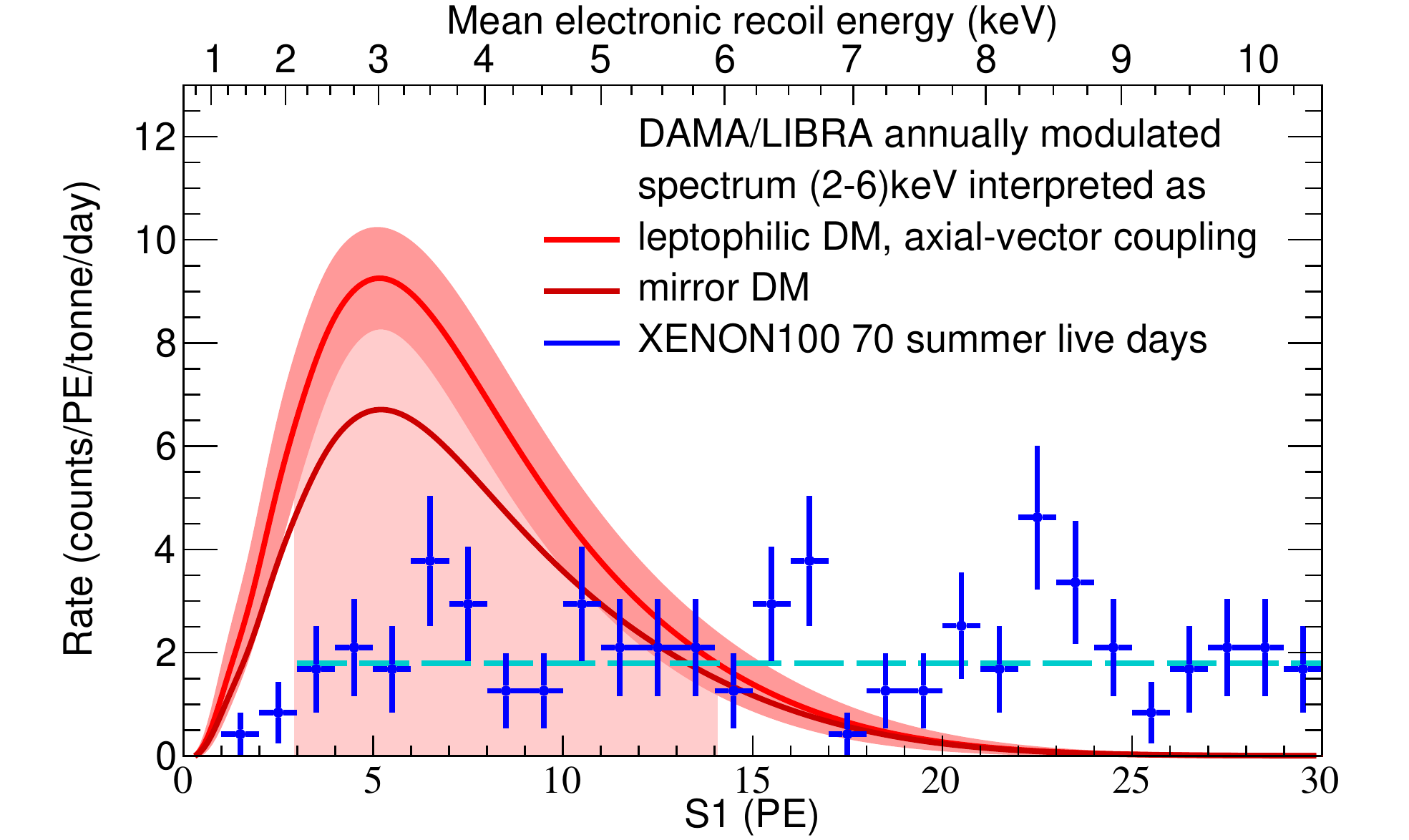}
\caption{Fig.~3. \textbf{Contrasting XENON100 data with DAMA/LIBRA}. The DAMA/LIBRA modulated spectrum (red), interpreted as WIMPs scattering through axial-vector interactions, as it would be seen in the XENON100 detector. The $1\sigma$ band includes statistical and systematic uncertainties. The DAMA/LIBRA modulated spectrum interpreted as luminous dark matter is very similar, whereas the interpretation as mirror dark matter is indicated separately (dark red). The (blue) data points are XENON100 data from the 70~summer live days with their statistical uncertainty. The expected average XENON100 rate is also shown (dashed cyan). The shaded region from (3--14)~PE was used to quantify the confidence level of exclusion.}
\end{figure}

The converted DAMA/LIBRA and measured XENON100 energy spectra are shown in Fig.~3. Part of the DAMA/LIBRA signal is expected to be seen below $2\1{keV}$ due to the finite energy resolution of XENON100. The uncertainty in the converted signal includes both the statistical uncertainty in the original DAMA/LIBRA energy spectrum~\cite{Bernabei:2013xsa} as well as the uncertainties from our energy conversion. The electronic recoil cut acceptance, shown in~\cite{Aprile:2014eoa}, was applied to the converted DAMA/LIBRA spectrum. The uncertainty shown in the XENON100 data is statistical.

The energy region to determine the level of exclusion was chosen starting at the threshold of 3~PE~\cite{Aprile:2012nq} to the point where the DAMA/LIBRA signal falls below the expected average XENON100 rate (cyan in Fig.3, calculated using a flat spectrum background model and scaled for the live time of the data set), which is at 14~PE, corresponding to (2.0--5.9)~keV. Taking systematic uncertainties into account, a simple comparison of the integral counts in this energy interval excludes the DAMA/LIBRA signal as axial-vector coupling between WIMPs and electrons at $4.4\sigma$ significance level, even considering all events from the well-understood XENON100 background~\cite{Aprile:2011vb} as signal candidates. To be consistent with previous analyses~\cite{Aprile:2014eoa}, the same data selection cuts were applied. The exclusion remains unchanged if we only impose a minimum set of requirements, namely that events have a single scatter in the fiducial volume with a prompt~S1 and delayed~S2 signal in the correct energy range. Furthermore, the exclusion stays above $3\sigma$ confidence level even if we consider a $4.5\sigma$ downward deviation in the measured data points~\cite{Manalaysay:2009yq,Aprile:2012an,Baudis:2013cca} that are used to set the energy scale, or if we set the light yield in xenon to zero below $2.9\1{keV}$, in contradiction with direct measurement~\cite{Aprile:2012an,Baudis:2013cca}.

\begin{figure}[htbp]
\centering\includegraphics[width=1\columnwidth,clip=true,trim=0pt 0pt 0pt 0pt]{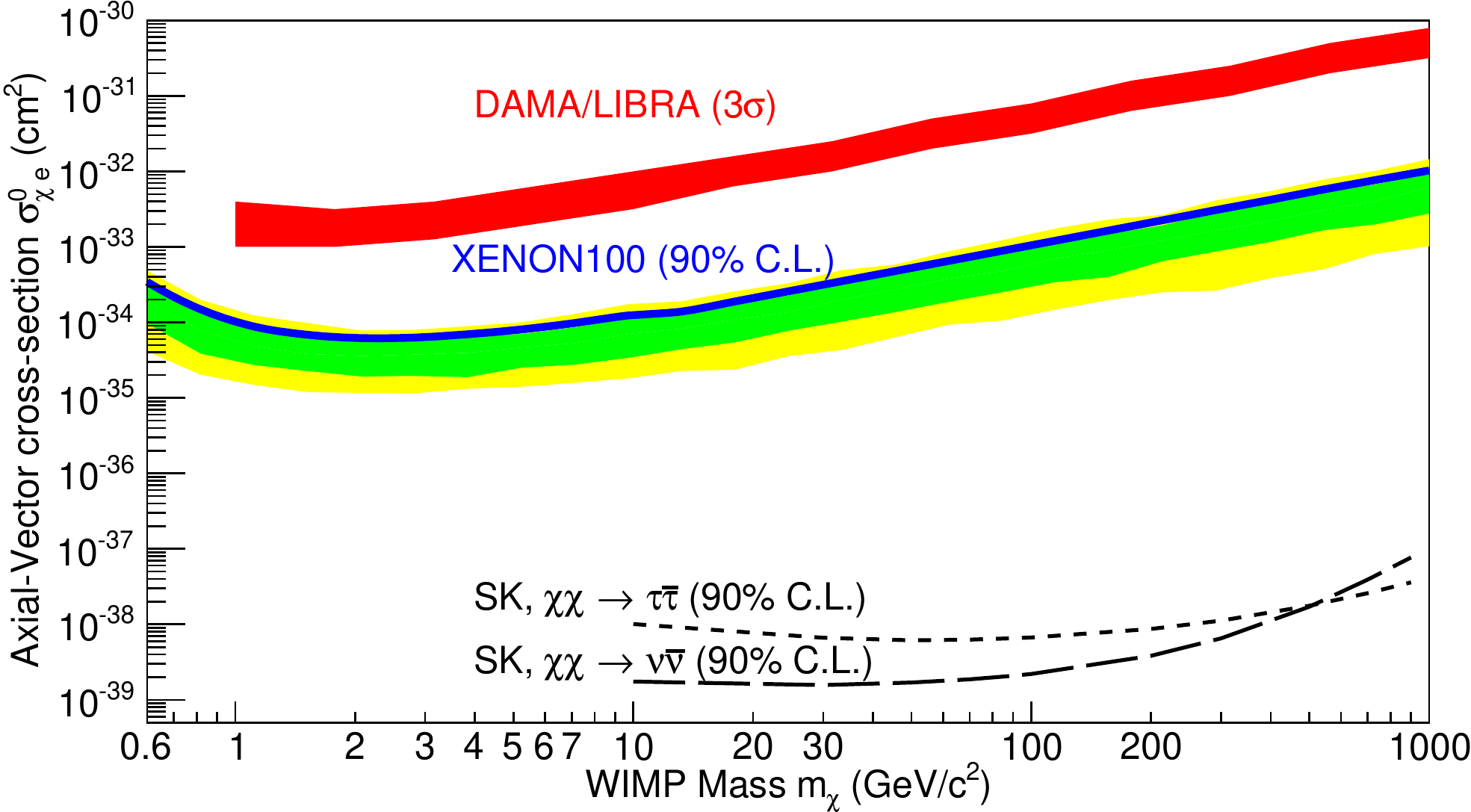}
\caption{Fig.~4. \textbf{Parameter space for WIMPs coupling to electrons through axial-vector interactions.} The XENON100 upper limit (90\% confidence level) is indicated by the blue line, along with the green/yellow bands indicating the 1$\sigma$/2$\sigma$ sensitivity. For comparison, we also show the DAMA/LIBRA allowed region (red) and the constraint from Super-Kamiokande (SK) using neutrinos from the Sun, by assuming dark matter annihilation into $\tau \bar{\tau}$ or $\nu \bar{\nu}$, both calculated in~\cite{Kopp:2009et}.}
\end{figure}

A profile likelihood analysis~\cite{Cowan:2010js,Aprile:2011hx} was performed to constrain the cross section $\sigma_{\chi e}^0 \equiv G^2 m_e^2/\pi $ for WIMPs coupling to electrons through axial-vector interactions. To this end, we drop the assumption of a 100\% modulated rate and use the entire $224.6$~live days data set. Fully analogous to~\cite{Aprile:2014eoa}, we use the same energy range and background likelihood function, derived from calibration data. We do not consider energy depositions below $1\1{keV}$, the lowest directly measured data point in~\cite{Aprile:2012an}. The resulting XENON100 exclusion limit (90\% confidence level) is shown (Fig.~4) along with the $1\sigma$/$2\sigma$-sensitivity bands based on the background-only hypothesis. It excludes cross-sections above $6\times10^{-35}\1{cm^2}$ for WIMPs with a mass of $m_\chi = 2\1{GeV/c^2}$. This is more than 5 orders of magnitude stronger than the one derived in~\cite{Kopp:2009et} based on data from the XENON10 detector, completely excludes the DAMA/LIBRA signal, and sets the strongest direct limit to date on the cross section of WIMPs coupling to electrons through axial-vector interactions.For comparison, we also show the DAMA/LIBRA allowed region and the constraint from Super-Kamiokande using neutrinos from the Sun, by assuming dark matter annihilation into $\tau \bar{\tau}$ or $\nu \bar{\nu}$, both calculated in~\cite{Kopp:2009et}. The XENON100 data completely excludes the DAMA/LIBRA signal and sets the strongest direct limit to date on the cross section of WIMPs coupling to electrons through axial-vector interactions, excluding cross-sections above $6\times10^{-35}\1{cm^2}$ for WIMPs with a mass of $m_\chi = 2\1{GeV/c^2}$.

\textbf{Kinematically Mixed Mirror Dark Matter:}
It has been suggested that multi-component models with light dark matter particles of $\sim$MeV/c$^2$ mass might explain the DAMA/LIBRA modulation~\cite{Foot:1991bp}. A specific example of such a model, kinematically mixed mirror dark matter~\cite{Foot:2014mia}, was shown to broadly have the right properties to explain the DAMA/LIBRA signal via dark matter-electron scattering. In this model, dark matter halos are composed of a multi-component plasma of mirror particles, each with the same mass as their standard model partners. The mirror sector is connected to the normal sector by kinetic mixing of photons and mirror photons at the level of $\sim10^{-9}$, which provides a production mechanism for mirror dark matter and a scattering channel with ordinary matter. While mirror hadrons would not induce nuclear recoils above threshold, mirror electrons ($m_e' = 511\1{keV/c^2}$) would have a velocity dispersion large enough to induce $\sim$keV electronic recoils.

The differential scattering rate of mirror electrons is proportional to $g N n_{e'}$, where $g$ is the number of loosely-bound electrons, assumed to be those with binding energy $<1$~keV~\cite{Foot:2014mia}, $N$ is the number of target atoms and $n_{e'}$ is the mirror electron density.The detector-dependent quantities are $N$ and~$g$. In order to compare DAMA/LIBRA directly with XENON100, we apply a constant scaling of $g_{\rm{Xe}}/g_{\rm{NaI}} \cdot N_{\rm{Xe}}/N_{\rm{NaI}} = 0.89$ to the DAMA/LIBRA spectrum and use the same procedure as in the case of axial-vector coupling: We again consider only the DAMA/LIBRA modulation signal, use the 70~summer live days, model scintillation in liquid xenon as described previously, and simply compare integral counts up to the point where the DAMA/LIBRA signal falls below the expected average XENON100 background data rate (at 13~PE), without background subtraction. This excludes the DAMA/LIBRA signal as kinematically mixed mirror dark matter at $3.6\sigma$ confidence level.

\textbf{Luminous Dark Matter:}
The third model we consider is Luminous Dark Matter~\cite{Feldstein:2010su}, featuring a dark matter particle with a $\sim$keV mass splitting between states connected by a magnetic dipole moment operator. The dark matter particle upscatters in the Earth and later de-excites, possibly within a detector, with the emission of a real photon. The experimental signature of this model is a mono-energetic line from the de-excitation photon. A mass splitting $\delta = 3.3$~keV provides a good fit to the DAMA/LIBRA signal~\cite{Feldstein:2010su} which would be explained as scattering of a real photon from the de-excitation of a $\sim$GeV/c$^2$ dark matter particle that is heavy enough to undergo upscattering, but light enough to evade detection in other direct searches.

This signature is independent of the target material; only the sensitive volume affects the induced event rate. As rates are typically given per unit detector mass, scaling to volume is inversely proportional to target density. We thus apply a constant scaling factor to the differential rate in DAMA/LIBRA which is the ratio of the target densities $\rho_{\rm{NaI}}/\rho_{\rm{Xe}} = 1.29$ in order to compare it to XENON100. Proceeding as in the previous two cases, we exclude the DAMA/LIBRA signal as luminous dark matter at $4.6\sigma$ confidence level. Together with the other two exclusions presented above, this robustly rules out leptophilic dark matter interactions as cause for the DAMA/LIBRA signal.

\textbf{Acknowledgements:} We thank J.~Kopp for providing the calculated wave functions and for useful discussions. We gratefully acknowledge support from the National Science Fundation, Department of Energy, Swiss National Science Foundation, Volkswagen Foundation, Bundesministerium fur Bildung und Forschung, Max Planck Gesellschaft, Research Center Elementary Forces and Mathematical Foundations, Foundation for Fundamental Research on Matter, Weizmann Institute of Science, Initial Training Network Invisibles, Fundacao para a Ciencia e a Tecnologia, Region des Pays de la Loire, Science and Technology Commission of Shanghai Municipality, National Natural Science Foundation of China, and Istituto Nazionale di Fisica Nucleare. We are grateful to Laboratori Nazionali del Gran Sasso for hosting and supporting the XENON project. XENON data is archived at the Laboratori Nazionali del Gran Sasso.

\textbf{Supporting Online Material:}
\begin{itemize}
\item S1 signal of low-energy electronic recoil events (\url{file1_s1energies.dat})
\item Recoil spectra for a 200GeV/c$^2$ WIMP with axial-vector couplings to electrons in NaI and Xe (\url{file2_axialvectorspectrum.dat})
\end{itemize}

\end{document}